\documentclass[11pt]{article}
\usepackage{amsmath}
\usepackage{amssymb}
\newcommand{\be}{\begin{equation}}
\newcommand{\ee}{\end{equation}}
\newcommand{\beq}{\begin{equation}}
\newcommand{\eeq}{\end{equation}}
\newcommand{\ba}{\begin{eqnarray}}
\newcommand{\ea}{\end{eqnarray}}

\begin{document}
\baselineskip=15.5pt
\pagestyle{plain}
\setcounter{page}{1}
%--------+---------+---------+---------+---------+---------+---------+
%Body

% Ofer's definitions

%\def\del{{\partial}}
%\def\vev#1{\left\langle #1 \right\rangle}
%\def\cn{{\cal N}}
%\def\co{{\cal O}}
%%\newfont{\Bbb}{msbm10 scaled 1200}     %instead of eusb10
%\newcommand{\mathbb}[1]{\mbox{\Bbb #1}}
%\def\IC{{\mathbb C}}
%\def\IR{{\mathbb R}}
%\def\IZ{{\mathbb Z}}
%\def\RP{{\bf RP}}
%\def\CP{{\bf CP}}
%\def\Poincare{{Poincar\'e }}
%\def\tr{{\rm tr}}
%\def\tp{{\tilde \Phi}}

%
%\def\TL{\hfil$\displaystyle{##}$}
%\def\TR{$\displaystyle{{}##}$\hfil}
%\def\TC{\hfil$\displaystyle{##}$\hfil}
%\def\TT{\hbox{##}}
%\def\HLINE{\noalign{\vskip1\jot}\hline\noalign{\vskip1\jot}}
%\def\seqalign#1#2{\vcenter{\openup1\jot
%  \halign{\strut #1\cr #2 \cr}}}
%\def\lbldef#1#2{\expandafter\gdef\csname #1\endcsname {#2}}
%\def\eqn#1#2{\lbldef{#1}{(\ref{#1})}%
%\begin{equation} #2 \label{#1} \end{equation}}
%\def\eqalign#1{\vcenter{\openup1\jot
%    \halign{\strut\span\TL & \span\TR\cr #1 \cr
%   }}}
%\def\eno#1{(\ref{#1})}
%\def\href#1#2{#2}
%\def\half{{1 \over 2}}

%--------+---------+---------+---------+---------+---------+---------+
%Hirosi's macros:
\def\ads{{\it AdS}}
\def\adsp{{\it AdS}$_{p+2}$}
\def\cft{{\it CFT}}

\newcommand{\ber}{\begin{eqnarray}}
\newcommand{\eer}{\end{eqnarray}}

\newcommand{\beqar}{\begin{eqnarray}}
\newcommand{\eeqar}{\end{eqnarray}}
\newcommand{\lm}{\lambda}\newcommand{\Lm}{\Lambda}
\newcommand{\ldm}{Lorentz/diffeomorphism}

%--------+---------+---------+---------+---------+---------+---------+

\newcommand{\nonu}{\nonumber}
\newcommand{\oh}{\displaystyle{\tfrac{1}{2}}}
\newcommand{\dsl}
  {\kern.06em\hbox{\raise.15ex\hbox{$/$}\kern-.56em\hbox{$\partial$}}}
\newcommand{\sqg}{{\sqrt{-g}}}
\newcommand{\gumn}{{g^{\mu \nu}}}
\newcommand{\gdmn}{{g_{\mu \nu}}}
\newcommand{\Gdmn}{{G_{\mu \nu}}}
\newcommand{\goo}{{g^{00}}}
\newcommand{\gdoo}{{g_{00}}}
\newcommand{\eeqarr}{\end{eqnarray}}
\newcommand{\pa}{\partial}
\newcommand{\gab}{{g^{\alpha \beta}}}
\newcommand{\Roo}{{R_{00}}}
\newcommand{\Rrr}{{R_{rr}}}
\newcommand{\Rthet}{{R_{\theta \theta}}}
\newcommand{\nupp}{{\nu^{\prime\prime}}}
\newcommand{\nup}{{\nu^{\prime}}}
\newcommand{\labpp}{{\lambda^{\prime\prime}}}
\newcommand{\labp}{{\lambda^{\prime}}}
\newcommand{\Ttoo}{{\tilde{T}_{00}}}
\newcommand{\Too}{{T_{00}}}
\newcommand{\Tij}{{T_{ij}}}
\newcommand{\Ttij}{{\tilde{T}_{ij}}}
\newcommand{\Toi}{{T_{0i}}}
\newcommand{\Ttoi}{{\tilde{T}_{0i}}}
\newcommand{\Tmn}{{T_{\mu \nu}}}
\newcommand{\Ttmn}{{\tilde{T}_{\mu \nu}}}
\newcommand{\ZZ}{{\rm \kern 0.275em Z \kern -0.92em Z}\;}

\begin{titlepage}

\begin{center} \Large \bf Space-times produced by a time-dependent scalar field\footnote{Talk presented at PASCOS Meeting (2005).}

\end{center}

\vskip 0.3truein
\begin{center}
 S.-Y. Pi

\vspace{0.3in}

Department of Physics \\
Boston University, Boston, MA 02215,
USA

\vspace{0.3in}

Center for Theoretical Physics, Massachusetts Institute of
Technology \\
Cambridge, MA 02139, USA
%Department of Physics \\
%Boston University, Boston, MA 02215,
%USA
\vspace{0.3in}

\end{center}
\vskip 1truein

\begin{center}
\bf ABSTRACT
\end{center}
Models of Lorentz/diffeomorphism violation frequently make use of a time-dependent scalar field. We investigate space-times produced by such a field.

\vskip2.6truecm
\leftline{BUHEP-05-10 }
\leftline{MIT-CPT/3676}
%\leftline{hep-th/0509037 \hfill date}
\smallskip
\end{titlepage}
\setcounter{footnote}{0}

These days there are extensive studies of field theoretic models that incorporate the possibility of  breaking CTP and/or \ldm\ invariance in gravity \cite{pi01}. Typically, these models contain either a vector field or a scalar field, whose vacuum expectation value selects the direction of symmetry breaking. Space-time symmetries are lost while spatial symmetries are preserved if, in a preferred frame, the vector field develops a time-like expectation value $v^\mu = (v, {\bf 0})$, with a constant $v$, or if the scalar field gets a time-dependent expectation value $\theta (t)$, and $v^\mu = \partial^\mu\, \theta$.

The self-consistently determined space-time produced by $v^\mu = (v, {\bf 0})$ or $\theta (t)$ through Einstein equation would provide a background about which gravity theory should be expanded. Then \ldm\ symmetry would be broken in the resulting theory. However, the above form of expectation values often requires an exotic or rather  unnatural form for the vector or scalar field Lagrangian in these models \cite{pi01}.

In this talk I shall discuss my recent research, motivated by the above topic, which is carried out in collaboration with Roman Jackiw. Instead of considering a scalar field with unconventional dynamics, we adopt a toy model which consists of gravity and a minimally coupled massless scalar field $\theta$, and we study the space-times produced by a time-dependent solution $\theta (t)$. The field equations are Einstein equation  and the equation of motion for $\theta$,
\ba
\Gdmn &=& 8\pi\, G\ \Tmn \label{eq:1}\\
D^2 \theta &=& \tfrac{1}{\sqg}\ \pa_\mu\ (\sqg\ \gumn\, \pa_\nu \, \theta) = 0 \label{eq:2}
\ea 
where $\Gdmn = R_{\mu \nu} - \tfrac{1}{2}\ \gdmn \, R$ and $\Tmn$ is the energy-momentum tensor given by
\be
\Tmn = \pa_\mu \, \theta \pa_\nu \, \theta - \tfrac{1}{2}\ \gdmn\, \gab\, \pa_\alpha \, \theta\, \pa_\beta\, \theta. \label{eq:3}
\ee
Using (\ref{eq:3}), the Einstein equation may be written as 
\be
R_{\mu\nu} = 8 \pi\, G \ \pa_\mu \, \theta\, \pa_\nu\, \theta. 
\label{eq:4}
\ee

We considered two kinds of solutions: (i) a spherically symmetric, time-dependent metric which we call a ``vacuum" configuration; (ii) a Robertson-Walker metric which may be called a ``cosmological" solution. For both cases, only the diagonal components of $R_{\mu\nu}$ are non-vanishing. The space-time component of Einstein's equation (\ref{eq:4}),
\be
R_{0 i} = 0 = 8\pi\, G\, \dot{\theta}\, \pa_i\, \theta ,
\label{eq:5}
\ee
requires either $\dot{\theta} = 0$ or  $\pa_i\, \theta= 0$. We posit the latter eventuality so that $\theta$ depends only on time. Then the remaining non-diagonal components $R_{ij}, i \ne j$, lead to vacuous equations. The diagonal components of $R_{\mu\nu}$ provide three differential equations:
\begin{subequations}\label{eq:6}
\ba
\Roo &=& 8\pi\, G \, \dot{\theta}^2 \label{eq:6a}\\
\Rrr &=& 0 \label{eq:6b}\\
\Rthet &=& 0. \label{eq:6c}
\ea
\end{subequations}
$R_{\phi \phi} = \sin^2 \theta\, \Rthet$ does not provide a new equation. For $\theta$, which depends only on $t$, the equation of motion (\ref{eq:3}) becomes
\be
\pa_0 (\sqg\ \goo \, \dot{\theta}) = 0.
\label{eq:7}
\ee
The above four equations, (\ref{eq:6a}, \ref{eq:6b}, \ref{eq:6c}) and (\ref{eq:7}) are key equations for both vacuum and cosmological solutions.

\subsection*{Vacuum Solution}
The most general form of the line element of a spherically symmetric, time-independent metric may be parameterized as
\be
d s^2 = e^\nu\, d t^2 - e^\lambda\, d r^2 - r^2\, d \Omega^2,
\label{eq:8}
\ee
with $\lambda$ and $\nu$ functions of only $r$ and $d\Omega^2 = d\theta^2 + \sin^2\, \theta\,  d\phi^2$. For this metric, the solution to (\ref{eq:7}) is given by
\be
\theta (t) = v t,
\label{eq:9}
\ee
where $v$ is an arbitrary constant. Then equations (\ref{eq:6a})-(\ref{eq:6c}) lead to the following differential equations.
\begin{subequations}\label{eq:10}
\ba
\tfrac{\nup}{r} + \tfrac{1}{2}\ (\nupp +\tfrac{1}{2}\ \nup^2 - \tfrac{1}{2} \labp \nup ) &=& 8\pi \, G\, v^2\, e^{(\lambda - \nu)} \label{eq:10a}\\
\tfrac{\labp}{r} - \tfrac{1}{2}\ (\nupp + \tfrac{1}{2}\ \nup^2 - \tfrac{1}{2}\ \labp\, \nup ) &=& 0 \label{eq:10b}\\
\nup - \labp = \tfrac{2}{r} \ (e^\lambda - 1) \label{eq:10c} \hspace{.5in}
\ea
\end{subequations}
(Prime denotes differentiation with respect to $r$.) Expressing $\nup$ in terms of $\lambda$ using (\ref{eq:10c}), (\ref{eq:10b}) may be written as
\be
\labpp + \tfrac{3\labp}{r}\ (e^\lambda - 1) + \tfrac{2}{r^2}\ (e^\lambda - 1) (e^\lambda - 2) = 0,
\label{eq:11}
\ee
while the sum of (\ref{eq:10a}) and (\ref{eq:10b}) becomes
\be
e^{-\nu} = \tfrac{2}{\mu^2r^2}\ [1+(r\labp - 1) \, e^{-\lambda}],
\label{eq:12}
\ee
where $\mu^2 \equiv 8\pi\ G v^2$. Eq. (\ref{eq:11}) possesses a spurious solution of the system, $e^\lambda = r/r-c$. For $c > 0$ this is the Schwarzchild solution which satisfies, (\ref{eq:10b}) and (\ref{eq:10c}), but does not satisfy (\ref{eq:10a}) where $\mu^2 \ne 0$. There are two self-evident solutions of (\ref{eq:11}): $e^\lambda =1$ and $e^\lambda =2$.  However, $e^\lambda = 1$ is a special case of the spurious solution with $c = 0$. The solution $e^\lambda = 2$ gives $e^\nu = \mu^2 r^2$ leading to the line-element \cite{pi02}
\be
d s^2 = \mu^2\, r^2\, d t^2 - 2 d r^2\, - r^2\, d \Omega^2.
\label{eq:13}
\ee

It appears that all solutions tend to the above expression at large distances. This follows from an analysis in which it is assumed that the asymptotic expression can be expanded in dominant and subdominant terms. We find,
\begin{subequations}
\ba
\lambda &=& l n\, 2 + \tfrac{\alpha}{\mu r} \ \cos\ (\sqrt{3}\ l n\, \mu r\, + \beta) + ...\label{eq:14a}\\
\nu &=& 2 \ l n \, \mu r - \tfrac{\alpha}{\mu r}\  [\cos\ (\sqrt{3} \ l n\, \mu  r + \beta) + \sqrt{3} \ \sin\ (\sqrt{3}\ l n \, \mu r + \beta) ] \label{eq:14b}
\ea
\end{subequations}
where $\alpha$ and $\beta$ are arbitrary parameters. The next sub-leading terms do not introduce any additional parameters and behave as $(\mu r)^{-2}$ times trigonometric functions with twice the above arguments.

Thus a scalar field linear in time produces self-consistently a 2-parameter family of static, rotationally symmetric space-times, none of which is asymptotically flat. The only analytic solution we found, {\it i.e.} the expression in (\ref{eq:13}), is a special case which is independent of arbitrary parameters.

The system possesses scale invariance, $\lambda(r) \to \lambda(c r)$. Our explicit vacuum solution is the unique solution which is scale invariant, other solutions are scale covariant, in the sense that scale transformations change the two arbitrary parameters of the solutions. 

It is not apparent that the space-time described by the line-element (\ref{eq:13}) can be employed as a physically acceptable background about which gravity theory should be expanded. It possesses a singularity at $ r= 0$, which acts as an attractor for geodesics, whose paths can be determined as usual from the geodesic equation. On the plane where the polar angle is $  \pi /2$, the path has a simple form,
\ba
r (t) &=& \tfrac{r_0}{\cosh \omega t} \nonumber\\
\phi &=& \sqrt{2}\ \bar{\omega} t \label{eq:15}
\ea
where $\omega, \bar{\omega}$ and $r_0$ are constants of motion satisfying $\omega^2 +\bar{\omega}^{2} = \frac{\mu^2}{2}$. A particle starting out at $r_0$ spirals into the origin in infinite time.

\subsection*{Cosmological Solution}
For the Robertson-Walker metric,
\be
d s^2 = d t^2 - a^2(t)\ \big[ \tfrac{d r^2}{1- k r^2}\ + r^2 d \, \Omega^2\big] 
\label{eq:16}
\ee
where $k = \pm , 0$, the solution to (\ref{eq:7}) satisfies
\be
\dot{\theta} (t) = v \, a(t)^{-3}
\label{eq:17}
\ee
where $v$ is an arbitrary constant. The Einstein equations (\ref{eq:6}) read
\begin{subequations}\label{eq:18}
\ba
- 3 \tfrac{\ddot{a}}{a} = 8 \pi\, G \ v^2\, a^{-6} = \mu^2\, a^{-6} \label{eq:18a}\\
(a \ddot{a} + 2\dot{a}^2 + 2 k) \ \hat{g}_{ij} = 0
\label{eq:18b}
\ea
\end{subequations}
where $\hat{g}_{ij}$ is the metric for 3-dimensional comoving coordinates. The first integral of (\ref{eq:18a}) becomes
\be
\dot{a}^2 = \tfrac{1}{6}\ \mu^2\, a^{-4} - c
\label{eq:19}
\ee
where $c$ is an integration constant. Using (\ref{eq:18a}) and (\ref{eq:19}), one finds that $c=k$. The final integration of (\ref{eq:19}) involves elliptic functions for $k\ne 0$, so for simplicity we consider a flat Robertson-Walker metric, $k=0$.

We find that
\be
a(t) = (\tfrac{3}{2})^{1/6}\, (\mu t)^{1/3}.
\label{eq:20}
\ee
Substituting (\ref{eq:20}) into (\ref{eq:17}), we find
\be
\theta (t) = \tfrac{1}{\sqrt{12\pi\, G}}\ ln\, t.
\label{eq:21}
\ee
For this scenario, the universe expands as $t^{1/3}$, which is different from the expansion due to radiation or matter domination in the standard cosmology.

\subsection*{Hydrodynamic Formulation}
More information about the space-times produced by a homogenous, but time-dependent scalar field can be obtained from the hydrodynamic formulation of our equations. It is known that the energy-momentum tensor of a scalar field, that depends only on time, has an ideal fluid representation, provided $g_{0i}$ vanishes \cite{pi03}. If $g_{0i} = 0$, it follows that $\goo = \frac{1}{g_{00}}$. For both of our scenarios, the static spherically symmetric metric and Robertson-Walker metric, $g_{0i}$ vanishes. For our system, the energy-momentum tensor is from \cite{pi03},
\begin{subequations}\label{eq:22}
\ba
\Too &=& \dot{\theta}^2 - \tfrac{1}{2}\ \gdoo\ \goo\, \dot{\theta}^2 = \tfrac{1}{2}\ \dot{\theta}^2 \label{eq:22a}\\
\Tij &=& -\tfrac{1}{2}\ g_{ij}\ \goo\, \dot{\theta}^2 \label{eq:22b}\\
\Toi &=& 0. \label{eq:22c}
\ea
\end{subequations}
On the other hand, for the ideal fluid the energy-momentum tensor is of the form
\be
\Ttmn = - P\, \gdmn + (P +\rho)\ u_\mu u_\nu,
\label{eq:23}
\ee
where $P$ is the pressure, $\rho$ is the energy density and $u^\mu$  is the four-velocity of the fluid normalized to unity, $u^\mu u^\nu\, \gdmn = 1$. In the comoving coordinates where the fluid is at rest, $u^\mu = (1/\sqrt{\gdoo}, \, {\bf 0})$ and $u_\mu = (1/\sqrt{\gdoo}, \, {\bf 0})$, the energy momentum tensor is,
\begin{subequations}\label{eq:24}
\ba
\Ttoo &=& \gdoo\, \rho \label{eq:24a}\\
\Ttij &=& - g_{ij}\, P \label{eq:24b}\\
\Ttoi &=& 0. \label{eq:24c}
\ea
\end{subequations}
Comparison with (\ref{eq:22}) shows that
\be
\rho = P = \tfrac{1}{2}\ \goo\, \dot{\theta}^2. 
\label{eq:25}
\ee
A fluid with this equation of state is  called a ``stiff" fluid. For our two scenarios, we have
\be
\rho (r) = \frac{1}{16\pi\, G r^2} 
\label{eq:26} 
\ee
for the static, spherically symmetric space given in (\ref{eq:13}) and
\be
\rho (t) = \frac{1}{24\pi\, G t^2} 
\label{eq:27}
\ee
for the cosmological solution with $a(t) \sim t^{1/3}$. Note that (\ref{eq:26}) again shows the singular behavior at $r=0$.

The rather vast gravity-fluid mechanics literature , with an arbitrary equation of state, can be searched for our static solution. Only space-time (\ref{eq:13}) can be found there as a special case \cite{pi04}.
%--------+---------+---------+---------+---------+---------+---------+
%Body

\section*{Acknowlegements:}
This work is supported by funds provided by the U.S. Department of Energy (DOE) under cooperative research agreement \#DE-FG02-05ER41360. Hospitality at MIT-CTP during my sabbitical is appreciated.

\end{document}